# Numerical Modeling of Healthcare Materials

**Hervé Bulou*,**

IPCMS/UMR 7504 CNRS/Strasbourg University, Strasbourg, France, 0000-0002-3365-6394

*Corresponding Author full name, Qualifications

23 rue du Loess

F-67000 Strasbourg, France

herve.bulou@ipcms.unistra.fr

(+33) 388107095

## Abstract

Healthcare materials, whether they are natural or synthetic, are complex structures made up of simpler materials. Because of their intricate structure, composite materials are ideal for prosthetics because it is possible to tune their structure to get mechanical properties that are compatible with bone, thus encouraging biointegration. To be effective, implants must be properly suited to the host, which necessitates complete control over both the design of the implants as well as their evolution over time as they are used. In the case of composite implants, this means that, while material control at the macroscopic scale is required during implant shaping, the quality of the interface, which is determined by phenomena acting at the nanoscale, is also required. In this work, we show that such an issue can be resolved by resorting to a multi-scale approach and that, in this context, numerical modeling is a useful tool. Then we describe the principle of the process for investigating composite biomaterials using numerical modeling. Then, we give a concrete example of the protocol by talking about the first step of the "grafting from" method that Professors H. Palkowski and A. Carrado's teams came up with to make implants from hybrid biomaterials.

## I - Introduction

### The need to develop implants

One of the amazing characteristics of bone is its capacity for self-repair, which enables it to regenerate after injury. Realigning and maintaining the limb after a fracture may be adequate for healing since osteogenesis fills the gap left by the fracture by producing new tissue, restoring the bone's functional efficiency [1]. But occasionally, mechanical or biological issues stop the bone from healing on its own. Osteogenesis may not be sufficient to replace considerable bone substance losses caused by some diseases or surgical procedures (removal of tumors, cysts, or infection foci), necessitating the deployment of a support to aid in bone repair. Even though bone grafts are still now regarded as the gold standard for replacing missing bones, they are not without drawbacks, including the danger of infection, high cost, scarcity, resorption risk, and difficulty restoring complicated missing bone components [2-4]. Artificial material-based patient-specific implants (PSI) are a desirable solution to this issue. Implants must be biocompatible and biotolerant to be accepted [5-7]. Polymers, such as polymethyl methacrylate (PMMA), polyether ether ketone (PEEK), bioceramics (calcium phosphate, hydroxyapatite, etc.), and some metals are the primary materials that are compatible with human fluids and tissues [2,8,9]. Due to the fact that metal-based biomaterials represent about 70% of implanted surgical devices, metals play a significant role in medical technology. Because of its biocompatibility, non-corrosiveness, structural stability, and formability, titanium is one of the metals that is most frequently utilized for PSI [10-14]. A major issue with titanium and metals in general is the mechanical difference (stiffness) between the implant and the bone [15]. When a pure titanium-based implant is placed, the load that would normally be supported only by the bone is divided between the bone and the implant, decreasing the stress on the bone [16]. This "stress shielding of the bone" phenomenon brought on by implant insertion is harmful. Wolff's law [17] states that the bone structure develops in opposition to the strain placed on it. As a result, portions of the bone that are heavily loaded will have increased bone mass, but sections that aren't as strained may see a drop in bone mass [18]. This loss of bone mass is known as bone resorption, and it can cause the implant to become loose or even shatter [19]. Like metals, popular polymers like PMMA and PEEK have problems with bone's mechanical compatibility [20]. Although bioceramics have great biocompatibility, their high price and fragility are a significant barrier to their use as implants [21].

**The interest of hybrid biomaterials for the development of implants**

Hybrid biomaterials are an interesting way to get around the drawbacks of monomaterials. Depending on the style of construction, these are intricate structures created by combining various types of materials, carrying the benefits of their individual components while reducing their drawbacks. A hybrid sandwich biomaterial (SM) [22] made of Ti as skin sheets and PMMA or copolymers (PMMA-co-PBMA) (PBMA: poly(butyl methacrylate)) as core layers was recently developed in a Franco-German partnership by the teams of Prof. H. Palkowski in Clausthal and Prof. A. Carrado in Strasbourg.

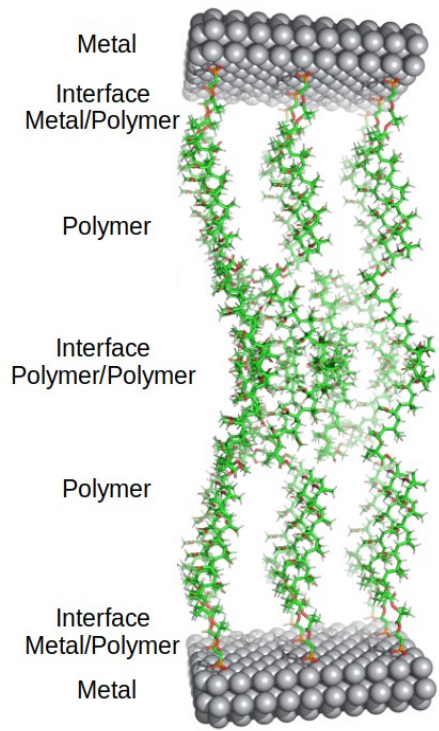

*Figure 1: The SM Implant schematic created by Professors H. Palkowski and A. Carrado's teams [28].*

The device developed by the two teams is shown schematically in Figure 1. We can make out the two metal plates at the ends that are separated by a number of interlocking polymer layers. The development of a method to achieve solid adhesion between the metal and the polymer is one of the significant contributions of the work. Epoxy resins are typically employed in SMs as an adhesive agent for aerospace and automotive applications [23]. They are unfortunately useless for biomedical applications due to their cytotoxicity [24]. The "grafting from" method was suggested and created by the teams of Prof. Palkowski and Carrado, and it entails grafting a polymer with the function of an adhesive agent onto titanium to ensure a covalent bond between the metal and the central polymer in the SMs. This method increases the lifetime and stability of the device while maintaining its non-toxicity [24-27]. On previously activated Ti substrates, the team under the direction of Prof. Carrado has successfully applied a method that enables the development of hybrid structures made of thick, stable, and biocompatible PMMA [28]. Then, the group under Prof. Palkowski prepared the SMs by placing two prepared Ti sheets in close proximity to an intermediate polymer sheet, heating them to the proper temperature, pressing them to create SMs, and then shaping them.

The primary benefit of this strategy is the versatility of the end product: the metal component gives the created composite great corrosion and mechanical resistance, while the polymer sheet offers the qualities of lightweight, high damping effects, and thermal resistivity [29,30]. By adjusting the ratio between the skin layer and the core layer and calculating them using the straightforward formula of mixing rules, the necessary mechanical properties can be easily obtained [31].

**The need for numerical modeling in biomaterials research**

As of now, in vivo tests of the produced SMs' cytotoxicity have shown that they are biocompatible [28]. This does not, however, ensure the applicability of the development of efficient implants from SMs. The durability of the implants under normal use conditions, in particular, is an important factor that must be improved so that new surgeries are performed as infrequently as possible over time. The quality of the biointegration of the implants to limit rejections, the conformability of the SMs to develop implants perfectly adapted to the patients, or the robustness of the interfaces between the different materials involved are all parameters that contribute to the durability of the implants, and so they should be quantified. A thorough analysis of each of these parameters requires a significant amount of work and research. Reduced stresses at the implant-bone contact have been shown to improve implant biointegration [32].The hybrid nature of SMs is particularly interesting from this point of view because it allows a large number of assembly options just by changing the nature of the metal and/or the polymer. However, even if it is possible to carry out an a priori preselection of the potentially interesting metal-polymer couples, their number remains high.

Regarding formaibility, PMMA is brittle at room temperature, so in order to increase formability and prevent core cracking during processing, the forming process should be carried out at higher temperatures [33]. Because the basal PMMA and the grafted PMMA creep together to form an adhesive bond, it is important to consider the adhesion conditions at the interface at higher temperatures [34]. Then, as the bonding conditions and the polymer are extremely sensitive to temperature and strain rates, it is required to conduct forming tests at various temperatures and strain rates in order to find the optimal processing parameters for a specific polymer-core SM [34,35]. Once more, numerous studies must be conducted in order to obtain the most thorough overview feasible (temperatures, thicknesses, strain rates, etc.).

Finally, the durability of implants is greatly conditioned by the state of the numerous interfaces present in the devices. They constitute points of fragility that must be characterized in order to hope to control them and improve the life span of implants based on SMs. In addition to their static characterization, which is mostly based on how well the materials stick together, they should also be tested for their resistance to wear caused by the tiny movements between the implant and the bone next to it [36]. These movements can cause wear debris to form, which can interact in ways that aren't good for living tissue [37].

The use of hybrid materials for the development of large-scale implants is constrained by the amount of time and resources required for each of these activities. Additionally, it can be challenging to gather certain information through experimental investigations that is necessary for the creation of large-scale implants. These relate to the processed SM's stress distribution, adhesion circumstances, etc. Without this information, it is impossible to calculate the likelihood of implant failure, and the patient could have early implant failure [38]. It also has to do with how shape affects the internal organization of SMs. The skins and core of Ti/PMMA/Ti SMs are subjected to multiaxial deformation modes during the production of the PSI [39-41]. These modes include compression, tension, and shear loads as well as bending (in and out of plane) and thickness variations. Additionally, the skin and core move in relation to one another as they form [42].

Numerical modeling is a potentially interesting approach to overcome these constraints. These techniques can help to reduce the cost of determining the preparation conditions of SMs and the most suitable material pairs for ISP purposes by allowing, on the one hand, to substitute some experiments with numerical simulations and, on the other hand, to explore research avenues that are inaccessible or difficult to reach experimentally (ultra-short time dynamics, nanometer scale observation).

In the rest of this article, we show that the description of composite biomaterials needs to be approached according to a multi-scale approach and that, in this context, numerical modeling is an appropriate tool. We present the protocol for the implementation of numerical modeling for the study of composite biomaterials (part II). Finally, we present a concrete illustration of one of the steps of this protocol. This is the activation phase of a titanium surface, the first step in the "grafting from" procedure proposed and developed by the teams of Prof. Palkowski and Carrado (Part III), and decisive for quantifying the robustness of the metal/polymer interface.

## II - Implementation of modeling to simulate hybrid materials

The implementation of numerical modeling to simulate hybrid materials implies that two particularly delicate aspects must be handled with attention in the modeling process. The first has to do with how the hybrid biomaterials' interfaces are constructed. They involve different interactions such as Pauli repulsion, electrostatic and covalent bonding, hydrogen bonds, and attractive Van der Waals dispersion forces [43]. As a result, the electronic distributions at the interface exhibit a very significant inhomogeneity that must be accurately characterized when attempting to gauge the cohesion of the interfaces. The vast range of spatial and temporal scales that must be taken into account when modeling hybrid biomaterial implants is the second challenging factor to handle. At the hybrid interface, there is usually an incommensurability between the metal and the organic layer, which leads to a very wide variety of environments for the molecules and the surface. This causes strain on the molecules as well as the surface [44].Such structural inhomogeneity can only be modeled by considering systems large enough to best describe the incommensurability between the metal and the organic layer.

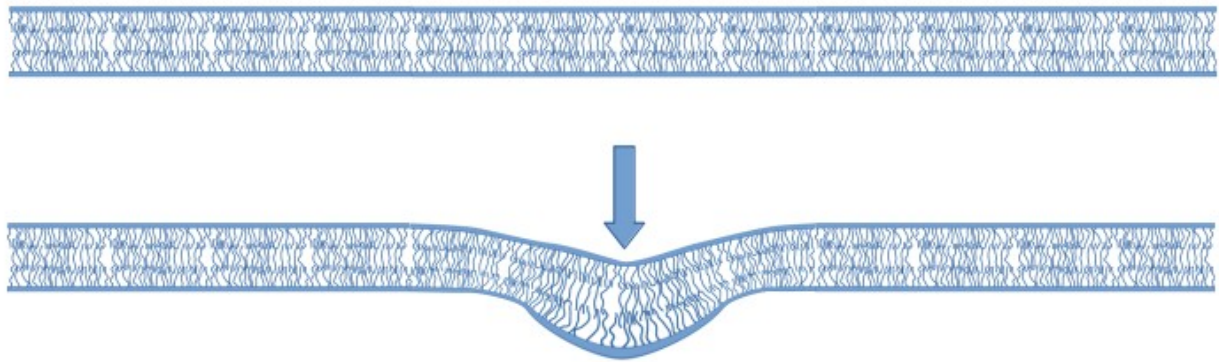

*Figure 2: The shape of the SM The SM, which is initially flat, experiences stress during the implant-shaping process. This stress can reflect at the metal-polymer interface and cause significant alterations to the interface that can cause the device to become brittle or even fail in the worst-case scenarios.*

In addition, the implant shaping process involves stamping the SMs. During this process, a macroscopic stress is applied at the interface of the SM with the external environment, as illustrated in Figure 2. However, the mechanical and elastic properties of these devices are governed at the nanoscale at the interfaces between the metals and polymers constituting the implant. This means that strain at the SM/external environment interface can induce significant deformations at the metal/polymer interfaces (Fig. 2). Therefore, in order to model the shaping of the implant, it is essential to ensure that a link between the phenomena acting at two very different scales is permanently established. The same applies to wear phenomena. In vivo implants are subjected to constant macroscopic stresses at the device/external environment interface.From a modeling point of view, these phenomena can be described as a succession of shocks to the implant over time. If the shocks undergone by the implant are at its interface with the external medium, their action is felt until the heart of the device, in particular at the metal/polymer interfaces, where the deformations propagate in the medium.

To represent the wear phenomenon, a significant temporal scale must be added to the huge spatial scale, taking into account both the succession of shocks on the implant and the propagation of mechanical waves within the device. There is currently no theory that can combine, within the same model, spatial scales with a resolution ranging from nanometers to meters and temporal scales with a resolution ranging from femtoseconds to months. Therefore, to model SM-based implants, it is necessary to use a panel of models, each coming from a different formalism and acting at different spatial and temporal scales, and used jointly by means of bilateral links established beforehand between the models. These links can take various forms, such as parameter sets adjusted for different models on the same data set, "back and forth" from one method to another during the simulation, etc. Different physical laws can be used to make the models. If electrons need to be taken into account, the models belong to quantum mechanics. If an atomic description of the matter is enough, the models belong to classical mechanics. If macroscopic systems need to be taken into account, the models belong to continuum mechanics.

For hybrid systems, the variety of possible environments for the atoms in the system (hybridization of atomic orbitals, atoms in non-equilibrium positions under high pressure, etc.) represents a real problem. On the one hand, this makes it difficult not only to evaluate the completeness of the configurations used to fit the force fields but also to determine simple but sufficiently general analytical forms to describe the diversity of environments. Our team has been working on an integrated strategy that combines quantum physics and classical physics techniques in this setting for almost ten years. The appropriateness of the structural parameters between the classical and quantum approaches is now ensured by a number of sets of parameters that have been derived using data collected from ab initio calculations performed on reference configurations. This is a necessary requirement for the coherence of the structural optimizations of the configurations carried out by combining the classical and quantum approaches. The traditional molecular dynamics code LAMMPS ("Large-scale Atomic/Molecular Massively Parallel Simulator") and our optimized force fields are compatible [45]. For really big systems, LAMMPS is more effective than "home" codes because it uses a domain decomposition method to distribute the tasks to be carried out on the many processors available. According to the type of bonds present in the system, we choose one of three force fields for structural pre-optimization: TBSMA [46-48], MEAM [49,50], or ABOP [51-53] for metallic substrates. Lennard-Jones potentials are used to simulate intramolecular interactions, while Universal Force Field potentials are used to simulate van der Waals interactions [54]. The "Bond-Order Morse" type potentials of Olmos-Asar, Rapallo, and Mariscal [55-56] describe interactions between molecules and metal substrates.

We carry out ab initio calculations within the framework of density functional theory (DFT) [58,59]. This theory allows for dealing with the shortest time and length scales (respectively, $10^{-12}$ s and $10^{-10}$ m). The DFT allows for the replacement of the solution of the multi-electron Schrödinger equation of a system of N interacting electrons by the solution of N eigenvalue equations (the Kohn and Sham equations), describing the behavior of a set of independent electrons evolving in an auto-coherent electrostatic potential determined from the distribution of electrons in the system. The atoms that make up matter are represented as central ions, which are treated as point positive charges and are surrounded by electrons, which are treated within the framework of quantum mechanics.DFT is based on the idea that the fundamental state of an interacting electronic system is entirely determined by the electron density of the system. By postulating a variational principle from this density [60], one can then determine the ground state of the system by minimizing the density functional associated with the total energy of the system [58].

There are several implementations of this theory. The most widely used is that of Kohn and Sham [59] which involves replacing a system of N interacting electrons by a fictitious system of N non-interacting electrons of the same density as the system of interacting electrons. The solution of the multi-electron Schrödinger equation of a system of N interacting electrons is then reduced to the solution of N eigenvalue equations (the Kohn and Sham equations), describing the behavior of a set of independent electrons evolving in a self-consistent electrostatic potential determined from the distribution of electrons in the system. Multi-electron exchange and correlation effects are taken into account by the electron density functionals governing the system. [61-63]. The Kohn and Sham electronic orbitals are represented in matrix form for the numerical solution using a representation of the Kohn and Sham eigenvalue equations built on a plane wave and/or Gaussian basis [64-66].

These orbitals and the energies that go along with them are determined by diagonalizing the Kohn and Sham matrix using either the direct diagonalization methods [67-69] or the iterative methods of Davidson [64,70-72]. The system's total energy is then calculated as the sum of the eigenenergies of the occupied Kohn and Sham orbitals, modified to account for multi-electronic effects [73]. Pseudopotentials are used to simulate the interactions between ions and electrons in the nucleus, which are thought of as classical point particles. The idea behind the pseudopotential approach is that the vast majority of a material's physical and chemical properties, including cohesion at interfaces, are solely dependent on the behavior of valence electrons. Thus, it may be said that the distribution of core electrons is mostly unaffected by the atom's chemical surroundings. Therefore, it is not necessarily required to explicitly take into account the core electron, especially when looking at the properties of materials. One of the key benefits of the pseudopotential approach is the about 60% reduction in the number of electrons that need to be handled for each atom when using solely valence electrons. As the core electrons are no longer explicitly involved in the calculations, it is no longer possible to describe the interaction of the valence electrons with the core ion via a simple Coulombic interaction. It is necessary to have an effective potential (the pseudopotential) produced by the inert ionic core (core plus core electrons). Several methods have been developed to define the pseudopotentials (conserved norm pseudopotentials [74,75], ultra-soft [76], projector-augmented waves [77], Goedecker-Teter-Hutter pseudopotentials [78]). The elementary excitations of matter, from which the properties of materials are derived, are described in the framework of density functional perturbation theory (DFPT), which allows one to determine the response of a quantum system, described in the DFT framework, to a small perturbation [79]. An important application of DFPT in the study of hybrid materials is the calculation of vibrational properties such as phonons, which can be used to calculate many physical properties such as infrared or Raman spectroscopy, specific heat, or thermal conduction, information that can then be confronted with experimental measurements. This information is also essential when one needs to model the impact of shocks acting on a material because it determines the propagation velocity of the deformation in the material.

## III - Results and Discussion

The complete modeling of an implant made of hybrid materials, such as the one developed by the teams of Professors H Palkowslki and A Carrado, is a consequential task whose complete presentation would significantly exceed the typical size recommended for this article. As a result, we have chosen to focus on the presentation of only a part of the modeling process of the implant rather than on the complete modeling of an implant made of hybrid materials. The first thing that has to be done in order to model hybrid biomaterial implants is to provide a description of the interfaces that make up the implant. The implants discussed in this article contain two different kinds of interfaces: the interface between the core PMMA and the grafted PMMA (shown as the polymer/polymer interface in Figure 1), which is formed by reptation between the core and grafted PMMA; and the interface between the metal and the grafted PMMA, which is shown as the metal/polymer interface in Figure 1. It is important to figure out how strong hybrid interfaces are because, in most cases, they are weak spots in composite materials. Only the characterization of the contact between the metal and the polymer is taken into consideration in this work. First, an alkaline solution is used to activate the Ti substrate. Then, a derivative of phosphonic acid is grafted onto the Ti substrate to make it functional. This is followed by a surface-initiated atom transfer radical polymerization using the phosphonic acid derivative as a coupling agent, a polymerization initiator, and malononitrile as a polymerization activator.

In this paper, we focus on the alkaline activation phase. During this phase, the titanium substrate is immersed in a sodium hydroxide solution, NaOH, at a temperature of 80 °C. We want to figure out the maximum density of PMMA molecules that can adhere to the titanium surface. The phosphonic acid derivative needs at least two hydroxyl groups to graft to the surface. So, we first made sure that it was possible to bind two hydroxyl groups at a distance that would allow the phosphonic acid derivative to be grafted. We investigated the adsorption of hydroxyl groups on the surface since the density of molecules adsorbed on the surface will depend on this activation phase. Three Ti(111) layers and 23 Å of vacuum were used to replicate the titanium substrate, and complete periodic boundary conditions were used to depict an infinite Ti(111) surface. The unit cell size in the direction parallel to the surface was 29.2×29.2 Å$^2$, and each layer comprises 10×10 titanium atoms.

Every configuration considered in this study underwent structural optimization. This takes up a significant amount of computing time. There are two ways to do this: a classical technique, which is less exact but faster, and a quantum mechanics-based approach, which is very accurate but expensive in terms of processing time. The comprehensive strategy outlined in the preceding section has been put into practice. Therefore, we started with a structural pre-optimization using the extremely quick "quenched" classical molecular dynamics [80], and then we moved on to a structural optimization using quantum mechanics. The Verlet algorithm is used to integrate the equations of motion in order to simulate the dynamics of the atoms in the system in classical "quenched" molecular dynamics [81]. The "quenching" procedure consists in examining for each atom at each integration step the relative orientation of the velocity of the atom with respect to the orientation of the force it is undergoing; when these two vectors are in opposite directions, the velocity of the atom is cancelled ("quenching"). In this way, the dynamics tends towards a system where all the atoms are in their most stable positions. The procedure for the structural optimization by DFT is the same as the one for the classical "quenched" molecular dynamics exposed above, except for the calculation of the forces experienced by the atoms, which is performed within the framework of the Hellmann-Feynman theorem [82,83] from the average value of the derivative of the total energy of the system with respect to the coordinates of the nuclei. The DFT calculations were performed with the CP2K program [84], which incorporates the Gaussian and plane waves (GPW) method [65,66]. The core electrons were described by Goedecker-Teter-Hutter pseudopotentials [78], while the valence electrons were expanded as a double-zeta Gaussian basis set. The atoms in the two higher planes of the substrate were not subject to any atomic positional constraints during the structural optimization phases. In contrast, the lower plane atoms were constrained to the locations that the titanium atoms occupy in terms of volume.

The adsorption energy, which represents the gain or loss of energy in the system associated with the adsorption of n hydroxyl groups on the titanium surface, was used to quantify the robustness of the adhesion. It is given by [85]

$$E_{ads} = \frac{E_{nOH/Ti(111)} - (E_{Ti(111)} - nE_{OH})}{n} \quad ,$$

where $E_{nOH/Ti(111)}$, $E_{Ti(111)}$, and $E_{OH}$ respectively represent the total energy of the system composed of n hydroxyl groups adsorbed on the surface of Ti(111), the total energy of the bare titanium substrate, and the total energy of an isolated hydroxyl group.

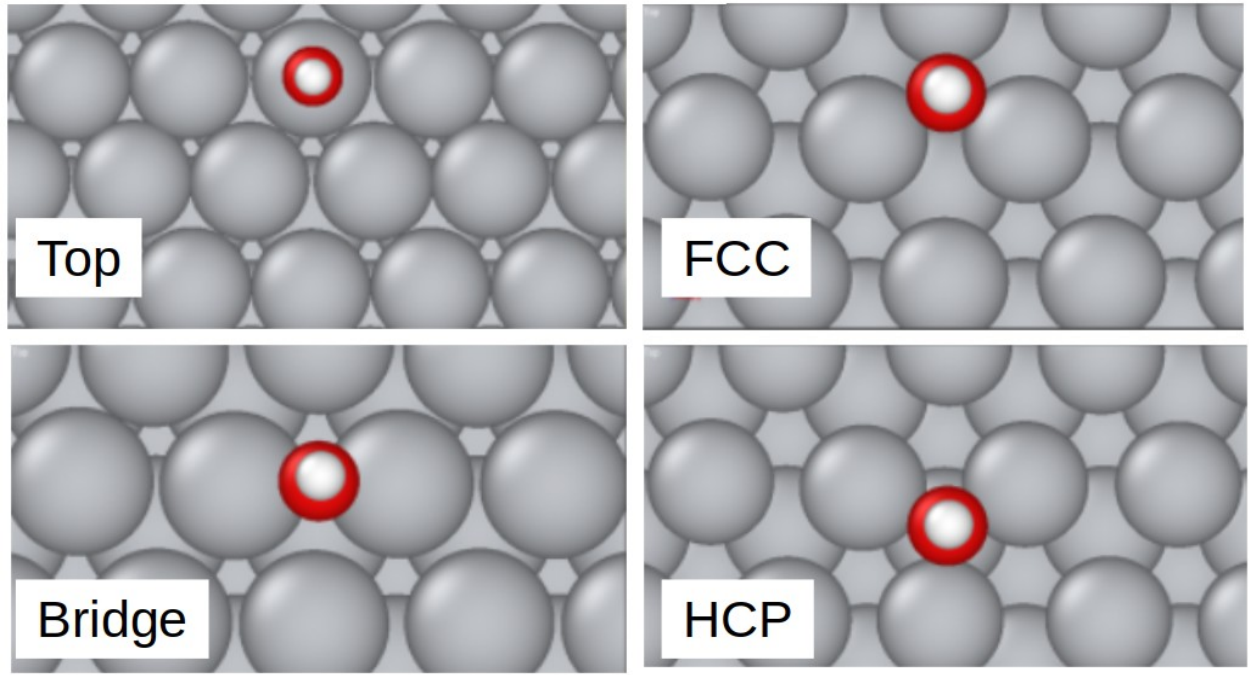

*Figure 3 shows the initial adsorption sites that were thought of for placing the hydroxyl group. Oxygen is shown in red, hydrogen is shown in white, and titanium is shown in grey.*

We have considered the (111) surface of a titanium substrate with a face-centered cubic structure. There are two categories of adsorption sites on this surface: HCP-type sites and FCC-type sites. For the study of hydroxyl group adsorption on Ti(111), we first put an hydroxyl group on each of these two sites. Then, we relaxed the structure of the system. We also considered two other initial positions: an on-top position and a bridge position (Figure 3). We observed two types of stable sites: HCP-type adsorption sites and FCC-type adsorption sites, with the FCC site being slightly more stable than the HCP site. On the other hand, when the hydroxyl group was initially placed in the bridge position or in the top position, we systematically observed a displacement of the hydroxyl group towards the HCP site or the FCC site. For the adsorption of the hydroxyl dimer, we started from the relaxed configuration of a hydroxyl group at the FCC site, and we positioned the second hydroxyl group in different initial sites around the first hydroxyl group. We have observed that the most stable position for the second hydroxyl group is also in the FCC position. The equilibrium distance between the two oxygen atoms is 3.04 Å, which is slightly greater than the distance between the titanium atoms (2.92 Å).

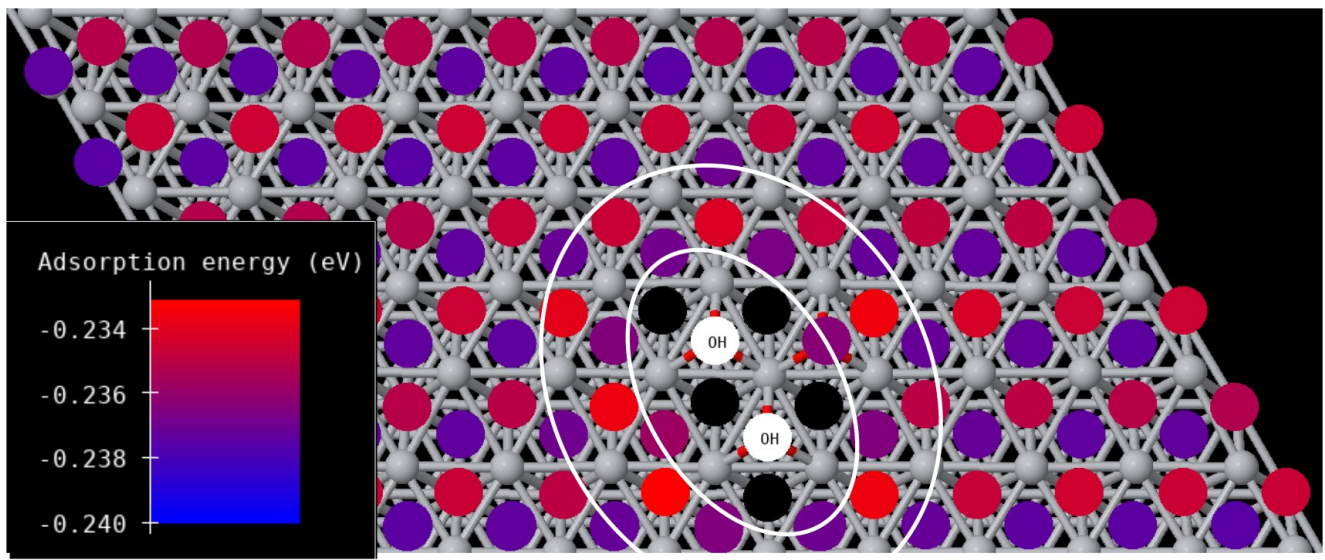

*Figure 4: Mapping of the adsorption energy of a configuration consisting of a hydroxyl dimer (symbol OH) and a hydroxyl monomer positioned at different positions on the surface. The areas in black correspond to unstable adsorption sites for the monomer. The ellipses represent the FCC adsorption sites located in the first and second crowns of the dimer.*

Figure 4 shows the adsorption energy map of a configuration consisting of a hydroxyl dimer (symbol OH) and a hydroxyl monomer positioned at different positions on the surface. It is observed that the HCP adsorption sites located in the immediate vicinity of the dimer are unstable. It is also observed that over the whole of the surface, the most stable adsorption sites for the monomer remain the FCC sites, as for the hydroxyls making up the dimer. Finally, it is observed that the FCC sites of the first two crowns around the dimer, although more stable than the HCP sites, are less stable than the most distant sites.

This adsorption-repulsion effect implies a limitation to the maximum density of hydroxyl groups (and consequently of phosphonic acid derivatives) that can be adsorbed on the titanium surface. Several phenomena can be at the origin of this repulsion effect (electrostatic effect, structural effect, etc.), and we are currently carrying out additional work to clarify this aspect of the study. It's important to figure out where this repulsion effect comes from so that you can reduce it and make the interface more stable.

## IV - Conclusions

In this article, we have shown that understanding the behavior of hybrid materials used to develop implants requires a multi-scale approach. In this context, we have also shown that numerical modeling of hybrid biomaterials is a way to solve this kind of problem. We then described a procedure for implementing numerical modeling for the study of hybrid biomaterials, and we presented an illustration of this implementation in the context of the formation of a metal/polymer interface. We have shown that this type of study provides important information that, among other things, allows us to quantify the robustness of the interfaces.

**ACKNOWLEDGMENTS**

This work was granted access to the HPC resources of IDRIS under the allocation A0120907459 made by GENCI. The authors would like to acknowledge the High Performance Computing Center of the University of Strasbourg for supporting this work by providing scientific support and access to computing resources. Part of the computing resources were funded by the Equipex Equip@Meso project (Programme Investissements d'Avenir) and the CPER Alsacalcul/Big Data. The autor is also grateful to P. Masson, A. Carrado, P. Masson, Z. Verrel and H. Iteney for stimulating discussions.